\documentclass[journal=jpcl,manuscript=article]{achemso}

\usepackage{chemformula} 
\usepackage[T1]{fontenc} 
\usepackage{physics} 
\usepackage{bm} 
\usepackage{hyperref}
 \hypersetup{
    colorlinks=true, 
    linkcolor=blue, 
    urlcolor=blue, 
    linktoc=all 
}
\usepackage{svg}
\usepackage{caption}

\author{Juan J. Aucar}
\affiliation[FCENA-UNNE]{Physics Department, Natural and Exact Science Faculty; National Northeastern University of Argentina, Avda Libertad 5460, W3404AAS, Corrientes, Argentina}
\alsoaffiliation[IMIT]{Institute for Modelling and Innovative Technology, IMIT (CONICET-UNNE), Avda Libertad 5460, W3404AAS, Corrientes, Argentina}

\author{Alessandro Stroppa}
\email{alessandro.stroppa@spin.cnr.it}
\affiliation{CNR-SPIN, c/o Dip.to di Scienze Fisiche e Chimiche - Via Vetoio - 67100 - Coppito (AQ), Italy}

\author{Gustavo A. Aucar} 
\affiliation[FCENA-UNNE]{Physics Department, Natural and Exact Science Faculty; National Northeastern University of Argentina, Avda Libertad 5460, W3404AAS, Corrientes, Argentina}
\alsoaffiliation[IMIT]{Institute for Modelling and Innovative Technology, IMIT (CONICET-UNNE), Avda Libertad 5460, W3404AAS, Corrientes, Argentina}
\email{gaaucar@conicet.gov.ar}

\title[An \textsf{achemso} demo]
  {On a Relationship Between the Molecular Parity-Violation Energy and the Electronic Chirality Measure}
\abbreviations{IR,NMR,UV}

\begin{document}

\begin{abstract}

When the weak-forces producing parity-violating effects are taken into account, there is a tiny energy difference between the total electronic energies of two enantiomers ($\Delta E_{PV}$), which might be the key to understand the evolution of the biological homochirality. We focus on the electronic chirality measure ($ECM$), a powerful descriptor based on the electronic charge density, for quantifying the chirality degree of a molecule, for a representative set of chiral molecules, together with their E$_{PV}$ energies.
Our results show a novel, strong and \textit{positive} correlation between $\Delta E_{PV}$ and $ECM$, supporting a subtle interplay between the weak-forces acting within the nuclei of a given molecule and its chirality. These findings suggest that experimental investigations for molecular parity violation detection should consider molecules with as large $ECM$ values as possible, and may support that a chiral signature is imprinted on life by fundamental physics via the parity-violating weak interactions.

\end{abstract}
\begin{figure}[H]\centering
\caption*{TOC Graphic.}
\includegraphics[height=5cm,width=5cm]
{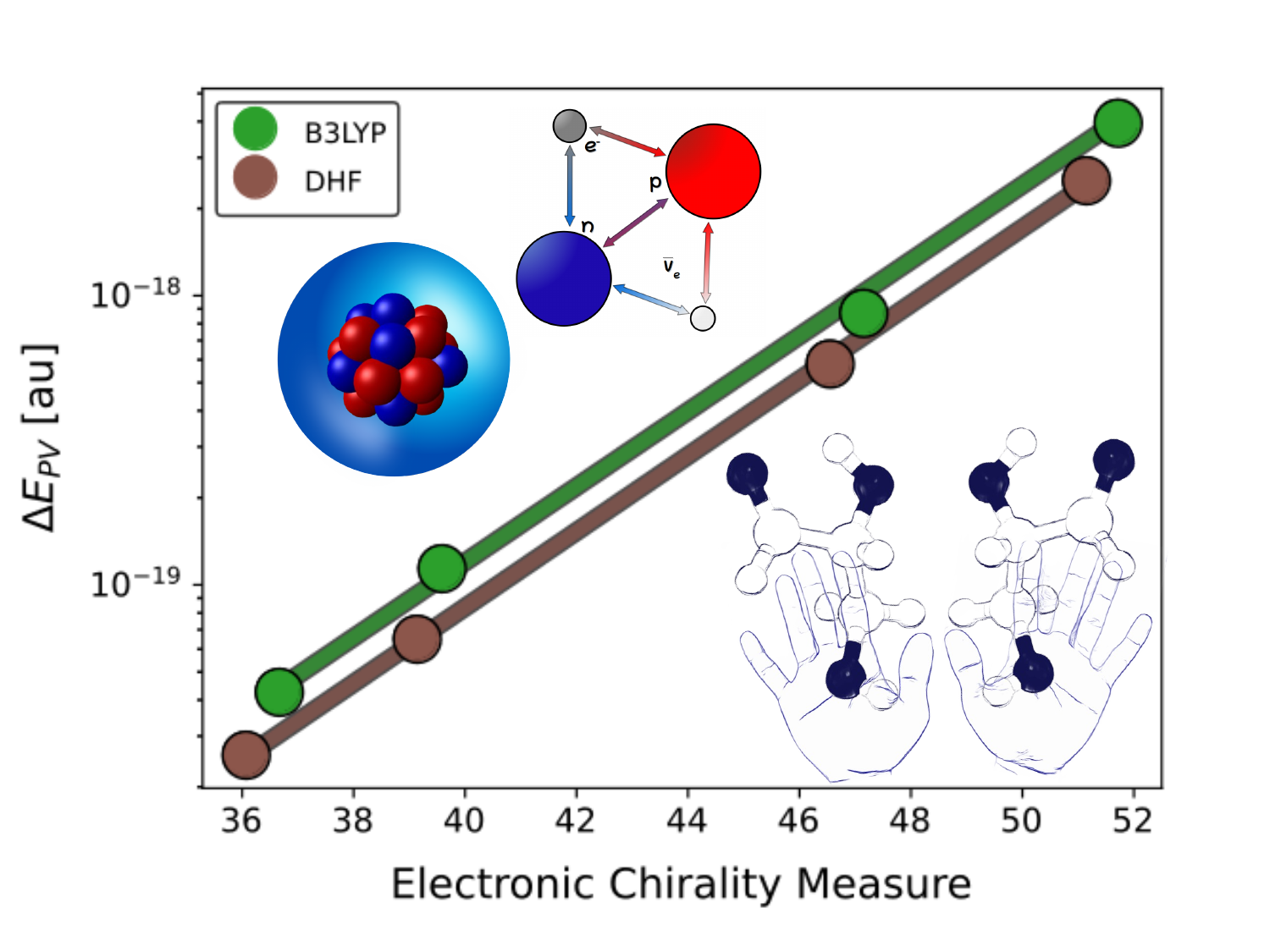}
\end{figure}
\section{Introduction}

Chirality is defined as `the geometric property of a rigid object (or spatial arrangement of points or atoms) of being non-superimposable on its mirror image; such an object has no symmetry operations of the second kind'.\cite{1997.Compendium,1997.IUPAC, kelvin1904} In three dimensions, every structure that lacks a mirror plane ($S_1$), an inversion center ($S_2$), or a higher improper rotation (roto-reflection) $S_n$ axis is chiral. The property of being `non-superimposable' means that the object and its mirror image are not directly congruent,\cite{2021.Nespolo} i.e. they cannot be mapped onto one another by an isometry of the first kind, or in other words by any combination of rotations and/or translations. A chiral object and its mirror image are called enantiomorphs or, for molecules, enantiomers. Two enantiomorphs are said to have opposite handedness (`left' and `right'), or chirality sense but they have similar physical and chemical properties. A sample in which all molecules have the same handedness is called enantiomerically pure, or enantiopure for short, or also said `homochiral' by some authors. Simple examples of chiral molecules are the hydrogen-peroxide (HOOH) and the ammonia isotopomer (NHDT).
\cite{2012.Naaman, 2020.Lu}

It is noteworthy that chirality plays an important role in the biological and pharmaceutical fields. \cite{1966.Cahn,2010.Kasprzyk}
Furthermore, chiral materials  possess many unique physical features, including circular dichroism, circularly polarized photoluminescence, nonlinear optics, ferroelectricity, extraordinary optical and spin-related properties.\cite{2021.Ma} For example, circularly polarized photoluminescence emission was found in chiral two-dimensional hybrid organic-inorganic perovskites, an important class of innovative materials, where 10\% polarization degree was detected in the absence of magnetic field. \cite{2018.Long, 2019.Ma,2019.Dong} \textit{Spin filtering} has been achieved with layers of chiral material via the so-called chirality-induced spin selectivity (CISS), in which the transport of one electron with one spin component is preferred over the other spin-component when propagating through the chiral material.\cite{2022.Evers}

From a more fundamental point of view, the origin of the molecular homochirality is still under debate.\cite{2022.Cowan} The following three hypothesis were proposed to explain it: the panspermia (life started somewhere in the universe and then arrived to the Earth), the primordial soup (in the early days of the Earth's evolution there was an appropriate atmosphere to promote the formation of basic organic compounds) and the theory of parity violation in weak interactions. \cite{2002_Quack, RBerger_Wires2018, 2021_Devinsky, MQuack_ChemSci2022}
The hypothesis that homochirality may arise from the weak-force acting on the nuclei of the molecular system has been put forward, among others, by several quantum chemists. \cite{RBerger_ChemPhysChem2000, JKLaerdahl_ChemPhysChem2000, 2011_Bast}
In relativistic quantum field theory, the CPT theorem states that the  combination of Time reversal (T), spatial inversion (P for parity), and  particle-antiparticle exchange (C for charge conjugation) must be a symmetry of any physical system. However, the discovery of parity-violating forces requires that a parity-violation (PV) interaction term must be taken into account, whenever the weak-forces are involved in the theoretical description of the physical system.

 The calculated energy differences between ground states of enantiomers due to parity-violating weak-forces are extremely small: about 100 aeV to 1 feV\footnote{1 atto-eV (aeV) = 10$^{-18}$ eV and 1 femto-eV (feV) = 10$^{-15}$ eV} (equivalent to 10$^{-11}$ to 10$^{-10}$ J mol$^{-1}$). Such a difference has not been detected experimentally yet. Its detection would have implications for fundamental problems in physics, in biochemistry and molecular biology; for example, the explanation of the evolution of `homochirality', referring to the  preference of one chiral form as building blocks in the biopolymers of all known forms of life (the L-amino acids in proteins and D-sugars in DNA, not the reverse D-amino acids or L-sugars).\cite{MQuack_ChemSci2022} In astrobiology, the spectroscopic detection of homochirality could be used as strong evidence for the existence of extra-terrestrial life, if any.

In the electroweak quantum chemistry it is known that the parity violation energy, E$_{PV}$, can be expressed as a sum of the contributions due to individual nuclei. Besides, for each nucleus, such interaction has a Z$_A$ high-power dependence, where Z$_A$ is the atomic number of nucleus A. Therefore, E$_{PV}$ will strongly depend on the contributions of the heaviest atoms, leading to a heavy-atom enhancement effect on E$_{PV}$.\cite{MQuack_ChemSci2022}
Great efforts have been put recently in order to quantify the chirality of a given molecule.\cite{2021_Barron,2009_Mislow} 
Zabrodsky and Avnir\cite{1995_Zabrodsky} proposed the continuous chirality measure ($CCM$), a `geometric' descriptor that depends on the distance of a molecular structure from its nearest achiral reference structure. This descriptor is definite positive, does not indicate handedness and has been used for describing asymmetric catalysis and protein backbones (vide infra).\cite{2019.Zahrt, 2018.Wang} In order to introduce handedness, a chirality measure must be a time-even pseudoscalar, i.e. sensitive to absolute configuration.\cite{2021_Barron} In line with previous findings, Hegstrom has suggested that the electron chirality density provides a natural convention for specifying the absolute chirality,\cite{1991_Hegstrom} but while the E$_{PV}$ strongly depends on the chirality of the electron density at distances very close to the nuclei,\cite{1999_Laerdahl} the chirality of the total electron density requires the integration of the density over the entire molecule, as recently reported by high-level calculations in amino acids.\cite{2020_Senami}

Following Grimme's approach,\cite{1998_Grimme} and based on Avnir's work\cite{1995_Zabrodsky}, Bellarosa {\it et al} realised that a similar approach could be based not only considering the geometry of the molecules, but also their electronic structures. They defined the descriptor known as electronic chirality measure ($ECM$), which was shown to have a good correlation with the enantiomeric excesses in a set of asymmetric aminohydroxylation reactions.\cite{2003_Bellarosa} The $ECM$ quantifies the chirality of a molecule by comparing its electronic wave function with the one corresponding to the nearest symmetric configuration. Based on its theoretical grounds it could carry some more subtle information related for example with the atomic number of the heavier atoms. This does not happens with other geometrical descriptors of chirality.

In this work, we focus on alanine, glyceraldehyde and several chiral organic ligands often used for the synthesis of chiral hybrid perovskites. By using the atomic charge of selected atoms as a parameter (say Z$_A$), we have being able to relate the results of relativistic and highly accurate calculations of $\Delta E_{PV}$ to  $ECM$ for the different chiral molecules in our selected set. Our study shows that in all the cases we have studied there exist a logarithmic correlation among E$_{PV}$ and the descriptor $ECM$. This finding clearly supports a subtle interplay between parity-violating interactions acting at the site of nuclei and the appearance of molecular chirality.


\section{Theoretical Approach}
In this Section we shall give a brief introduction about the calculation of the PV energy  in molecules within the relativistic framework. We will emphasise an important property, {\it i. e.} its $Z_A$ dependence, where $Z_A$ is the atomic number of the nucleus A in the molecule. Afterwards, we will describe the special descriptor of chirality used in this work, {\it i. e.} $ECM$ which also has a dependence with $Z_A$. For calculating it we have developed a software, which is here freely distributed.

\subsection{Calculation of parity-violation energies in molecules}
%
The parity-violating interactions occurring in molecules are mainly due to the exchange of virtual $Z^0$ neutral bosons, between the electrons, say $i$, and a given nucleus, say $A$. These interactions are described within a relativistic framework by the following Hamiltonian:\cite{Blundell_PRD1992}
\begin{equation}\label{eq:EPV-relativistic}
 H_{PV} =\frac{G_F}{2\sqrt{2}}\sum_{i,A}Q_{W,A}\gamma_i^5\rho_A(\textbf{r}_i)
\end{equation}
The constant $G_F$ is the Fermi electroweak coupling constant (G$_F$ $\approx$ 1/(300 GeV)$^2$), $\gamma^5$ is the Dirac-matrix which is proportional to the chirality operator and $\rho_A (\textbf{r}_i)$ is the normalized nuclear density. The parameter $Q_{W,A}$ is the weak charge of nucleus $A$, which is positive for all nuclei except hydrogen. It can be written as
\begin{equation}\label{eq:QWA}
 Q_{W, A} = -N_A + Z_A (1 - 4 \texttt{sin}^2 \theta_W)
\end{equation}
\noindent being $N_A$ the number of neutrons and $Z_A$ the number of protons. The angle $\theta_W$ is known as the Weinberg mixing angle and the Weinberg parameter, sin$^2 \theta_W$, slightly depends on the nuclear model used.

The Hamiltonian $H_{PV}$ of Eqn.\eqref{eq:EPV-relativistic} is the nuclear-spin-free PV operator written within a four-component (relativistic) framework. We have neglected other contributions, which  depend on the  nuclear-spin but they are less important in the computation of molecular energy.

\vspace{0.5cm}

To obtain the four-component parity-violation contribution to the electronic energy up to the first-order, one should calculate the  expectation value of the parity-violation Hamiltonian, $H_{PV}$,\citenum{Faglioni_2001}
\begin{equation}\label{eq:PVED_rel}
 E_{PV} =\expval{H_{PV} }{\Psi}
\end{equation}
Considering now that those PV nucleon-electron interactions occurs when the electrons are inside the nuclei. In such a case the parity violation energy shall be written as a sum of local contributions
\begin{equation}\label{eq:EPVA}
E_{PV} = \sum_{A=1}^N E_{PV}^A
\end{equation}
Furthermore, the PV effects originating in a nucleus A are known to be proportional to $G_F$ $\alpha$ $Z_A^3$ (being $\alpha$ the fine structure constant), and also to the relativistic spin-orbit effects which are proportional to $\alpha^2$ $Z_A^2$. Then, the order of magnitude of the PV contribution to the electronic energy given by a nucleus A is \cite{MQuack_ChemSci2022}
\begin{equation*}
    E^A_{PV} \approx G_F  \alpha^3 Z_A^5
\end{equation*}
Applying this approximation, we expect that the main energy contribution of Eqn.\eqref{eq:EPVA} will be proportional to the fifth power of the atomic charge of the heaviest atom of a given chiral molecule. A detailed description for the relativistic calculations of the PV energy in chiral molecules can be found in Ref. \citenum{1999_Laerdahl}.

As mentioned above, when PV weak interactions are considered, the ground state energies of enantiomers (usually expressed as R and S) of chiral molecules are slightly different. This energy difference is defined as
\begin{equation}\label{eq:PVED}
\Delta E_{PV} = 2E_{PV} = E^R_{PV} - E^S_{PV}
\end{equation}
%

%
\subsection{The electronic descriptor of chirality, ECM}
%
The $ECM$ descriptor gives an `electronic' measure of the chirality of any molecular system. The method defines an operator $\hat{S}=1-\hat{Z}$, where $\hat{Z}$ represent an operation that shifts the wave function centered at the original positions $p_i$ ($\Psi(p_i$)) to the same wave function centered at the atoms in new positions $\tilde{p}_i$ ($\Psi(\hat{p}_i)$) corresponding to the nearest symmetric achiral configuration of the molecule.\cite{2018_cosym} The total wave function is the same, though written on top of different geometric configurations: the whole set of molecular orbitals (MOs) $\ket{\phi_\alpha}$ of the chiral molecule and the $\ket{\widetilde{\phi}_\beta}$ MOs centered at the achiral molecular symmetry are connected by a rigid shift from the initial atomic configuration $p_i$ to the new configuration $\tilde{p}_i$.
In order to impose the orthonormality condition, the overlap matrices are introduced whose matrix elements are defined as $O_{ij}=\bra{\chi_i}\ket{\chi_j}$ and $O'_{ij}=\bra{\chi'_i}\ket{\chi'_j}$, where the atomic basis sets centered at positions $p_i$ and $\tilde{p}_i$ are ${\ket{\chi_i}}$ and ${\ket{\chi'_i}}$, respectively. The MOs are written as a linear combination of these atomic basis sets,

\begin{eqnarray}
    \ket{\phi_\alpha} &=& \sum C_{i \alpha} \ket{\chi_i}  \\
    \ket{\widetilde{\phi}_\beta} &=& \sum \widetilde{C}'_{j \beta}\ket{\chi'_j}
\end{eqnarray}
\noindent where $\widetilde{{\bf C}}'$ is the matrix whose elements
\begin{equation}\label{eq:Cachiralnewbasis}
\widetilde{C}'_{j \beta}=\bra{\chi_j'}\ket{\widetilde{\phi}_\beta}
\end{equation}
\noindent  represent the symmetric MOs in terms of ${\ket{\chi'_j}}$, and $C_{i\alpha}=\bra{\chi_i}\ket{\phi_\alpha}$. The wave function $\Psi$ is written as a determinant of occupied MOs.

Bellarosa et al. introduced a special procedure to generate the achiral orbitals at the new positions $\tilde{p}_j$. They are obtained from the transformation
\begin{equation}\label{eq:bellarosaequation}
\hat{Z}\Psi(p_i)=\Psi(\tilde{p}_i)=O'(\tilde{p}_i)^{-\frac{1}{2}}O(p_i)^{\frac{1}{2}}\Psi(p_i)
\end{equation}
From this last equation it is seen that\cite{2003_Bellarosa}
\begin{equation}\label{eq:ECM_tildeCnewbasis}
 \widetilde{C}'_{i\beta}=\left(O'^{-1/2}O^{1/2}C\right)_{i\beta}
\end{equation}
\noindent and so, the expectation value of $\hat{Z}$ can be written as
\begin{flalign}
 \expval{\hat{Z}}&= \bra{\Psi}\ket{\widetilde{\Psi}}\equiv\sum_\alpha^{occ}\sum_{i,j}^{naos}\bra{\chi_i}C_{i\alpha}\widetilde{C}'_{j\alpha}\ket{\chi'_j} \\
  &=\sum_\alpha^{occ}\sum_{i,j}^{naos}C_{i\alpha}\widetilde{C}'_{j\alpha}O_{ij}^M = \sum_\alpha^{occ}\left(C^\dagger O^M\widetilde{C}'\right)_{\alpha\alpha}
 \end{flalign}
where the elements of $\widetilde{{\bf C}}'$ are defined in Eqn. \eqref{eq:ECM_tildeCnewbasis}, $O^M_{ij}=\bra{\chi_i}\ket{\chi'_j} $ and $\expval{\hat{Z}} = \sum_{\alpha} \bra{\phi_\alpha} Z_{\alpha} \ket{\phi_\alpha}$.

In this way the descriptor $ECM$ can be obtained as
\begin{equation}\label{eq:ECM}
    ECM = \expval{\hat{S}}\times 100 \equiv 100\left(1-\left|\sum_\alpha^{occ}\left(C^\dagger O^M\widetilde{C}'\right)_{\alpha\alpha}\right|\right)
\end{equation}
There are two limiting cases to consider: a) $ECM$=0, meaning that the electronic wave function represents an achiral molecule, and b) $ECM$=100, meaning that both, the achiral and the chiral wave functions do not overlap each other. Furthermore, when the expectation value of $\hat{Z}$ is calculated in the chiral configuration the transformation projects out the whole of the chirality information. So, once this is done, a new iteration cannot \textit{extract} anything else.

\section{Computational Details}
We have focused on a selected set of molecules comprising alanine, glyceraldehyde and several selected some organic ligands  (in neutral form), which are often used in the synthesis of several chiral hybrid inorganic-organic perovksites (HIOPs).

The ionic and electronic optimization was performed by using DIRAC package \cite{DIRAC23} at Dirac-Hartree-Fock (DHF) and four-component (4C)-B3LYP \cite{B3LYP} framework with the dyall.cv2z \cite{dyall_cv2z} basis set. The theoretical approximations here are referred as DHF/dyall.cv2z and 4C-B3LYP/dyall.cv2z. The PV energy difference was calculated from Eqn. \eqref{eq:PVED} implemented in the DIRAC code.

The $ECM$ values for the different molecules were calculated according to Eqn. \eqref{eq:ECM} at both levels of theory, DHF/dyall.cv2z and 4C-B3LYP/dyall.cv2z with the recently implemented pyECM program package\cite{pyECM} that imports the wave function from the PySCF code \cite{2018_pyscf}. To the best of our knowledge, the implementation for the calculation of the descriptor $ECM$ in the full-relativistic framework is reported for the first time in this work.

The calculation of $ECM$ for each molecule proceed as follows:
\begin{enumerate}
   \vspace{-0.2cm} \item Geometry optimization;
    \vspace{-0.3cm}\item The achiral nearest symmetric structure is obtained through the
    Continous Symmetry Measures (CoSyM) website;\cite{2018_cosym}
    \vspace{-0.3cm}\item The $ECM$ is obtained according to Eqn. \eqref{eq:ECM} by using our pyECM code.
\end{enumerate}
We have developed the code pyECM whose main characteristics are given in the Supplementary Information.

\section{Results and Discussion}
%
We have calculated the energy difference $\Delta E_{PV}$ and the descriptor $ECM$ for different families of chiral molecules, whose geometries are given in the ESI file. We focused ourselves in the Z$_A$ dependence of both quantities by substituting one or two of their atoms. After the substitutions we re-optimized the geometries and then calculated the $\Delta E_{PV}$ and $ECM$.
For example, for the BPEAR molecule ((R)-1-(4-bromophenyl)ethan-1-amine) ) and the CHEAR molecule ((R)-1-cyclohexylethan-1-amine) the halogen atom or the nitrogen atom (marked with a cross in Figs. 1 and 2 of the ESI file) were substituted with heavier halogen atoms or other atoms belonging to the group XV of the Periodic Table, i.e. the nitrogen family which includes the following atoms: nitrogen (N), phosphorous (P), arsenic (As), antimony (Sb) and bismuth (Bi). In the case of the alanine two atoms were considered (see Fig. 3 of ESI). The same procedure was performed for the glyceraldehyde molecules. In Table \ref{Table:systems} the studied molecules, their acronyms and the $Z_A$ values of the substituent atoms are listed.

\begin{figure}[]
  \begin{minipage}[b]{0.5\linewidth}
    \centering
    \includegraphics[width=1.0\linewidth]{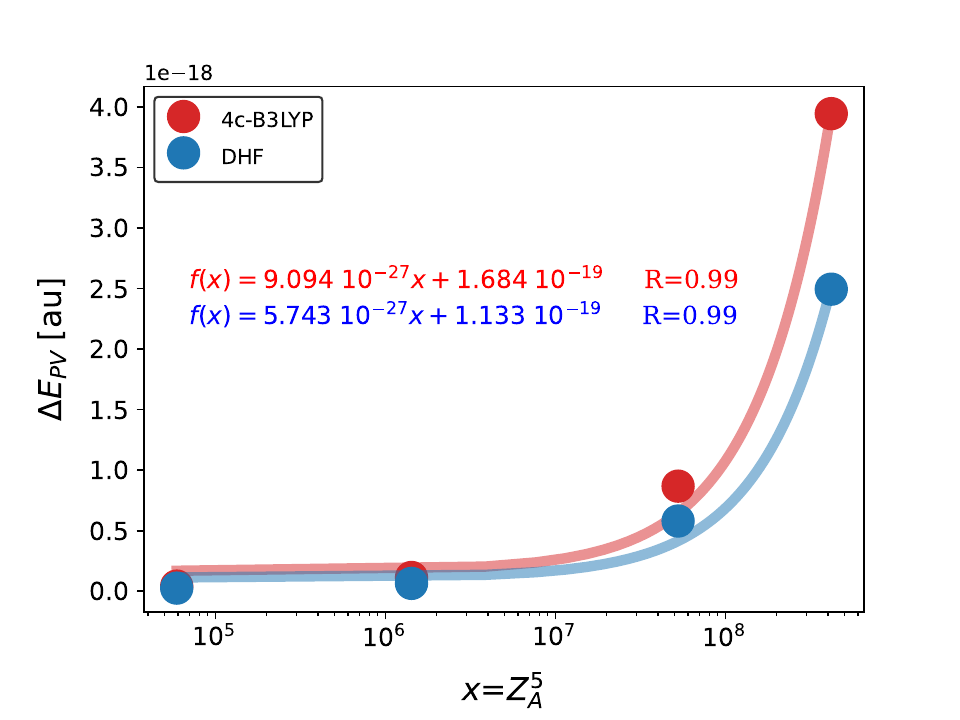}
  \end{minipage} \\
\begin{minipage}[b]{0.5\linewidth}
    \caption*{a)}
\end{minipage}
  \begin{minipage}[b]{0.5\linewidth}
    \centering
    \includegraphics[width=1.0\linewidth]
    {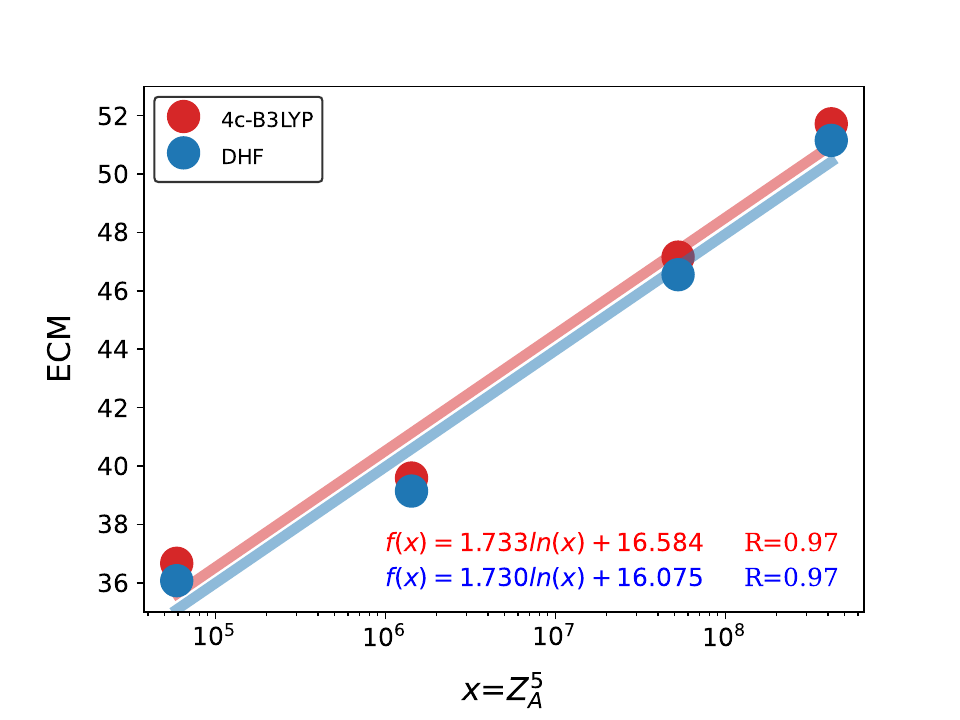}
    \caption*{b)}
  \end{minipage}
  \begin{minipage}[b]{0.5\linewidth}
    \centering
    \includegraphics[width=1.0\linewidth]{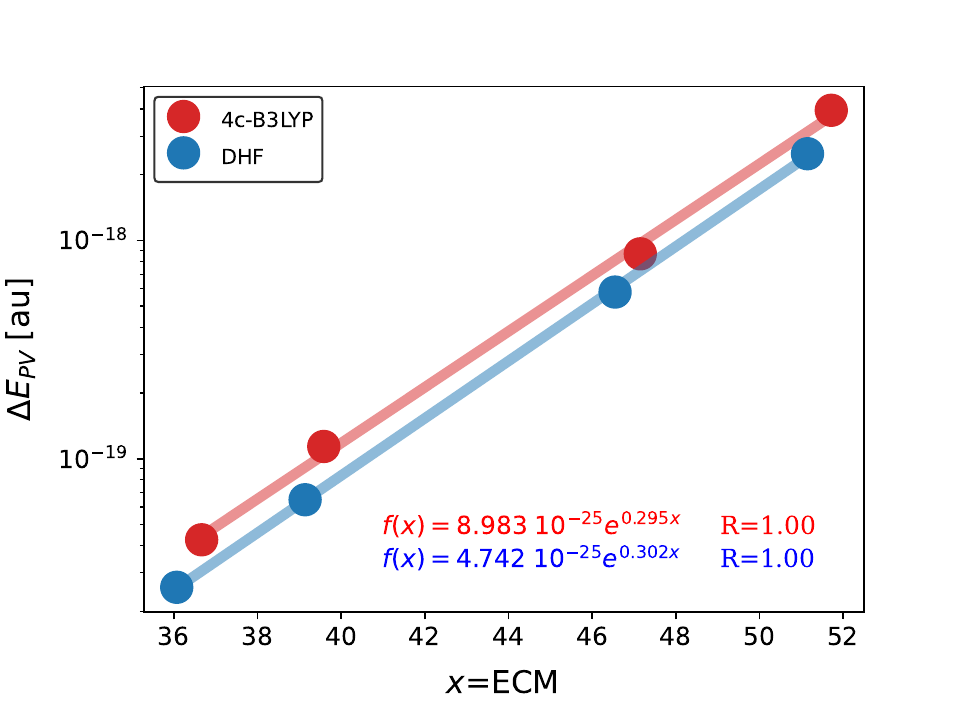}
    \caption*{c)}
  \end{minipage}
  \caption{
  Correlations of $\Delta E_{PV}$, $Z_A^5$ and $ECM$ for the BPEAR molecule. The substituted $A$ atom is shown in the ESI file. All calculations were performed at DHF/dyall.cv2z and 4C-B3LYP/dyall.cv2z levels of theory.
  }
  \label{fig:REL_BPEAR_cv2z_all}
\end{figure}

\begin{table}[]
\begin{tabular}{ccc}
\hline
\textbf{Molecule}                           & \textbf{Acronym}        & $Z_A$\textbf{ values} \\ \hline
Alanine                            & alanine        & (8),16,34,52         \\ \hline
Glyceraldehyde                     & glyceraldehyde & (8),16,34            \\ \hline
(R)-1-(4-bromophenyl)ethan-1-amine & BPEAR          & 9,17,(35),53         \\ \hline
(R)-1-cyclohexylethan-1-amine      & CHEAR          & (7),15,33            \\ \hline
\end{tabular}
\caption{The molecules studied, their acronyms and the atomic numbers, $Z_A$, of the substituted atoms. Between parenthesis there are the original $Z_A$}.
\label{Table:systems}
\end{table}
\vspace{0.3cm}
To start with, we focus on the BPEAR molecule. In Fig. \ref{fig:REL_BPEAR_cv2z_all} a) we show the dependence of $\Delta E_{PV}$ with the fifth power of $Z_A$, i.e. the nuclear charge of the atom being replaced by others which are members of the same column of the Periodic Table. We observe a clear relation that holds very similar for both kind of calculations, meaning correlated (4C-B3LYP) and uncorrelated (DHF). The slope remains of the same order of magnitude in both cases.
Then one may search for a similar correlation between $ECM$ and $Z_A$, meaning a dependence of $ECM$ with $Z_A^5$. In Fig. \ref{fig:REL_BPEAR_cv2z_all} b) we show the relation we found among $ECM$ and $Z_A^5$, where the correlation between $ECM$ and $Z_A^5$ is logarithmic. Again the electron correlation effects are negligibly small.

These findings suggest that there may be a relation between $\Delta E_{PV}$ and the descriptor $ECM$. In fact, we found an exponential relation that is shown in Fig. \ref{fig:REL_BPEAR_cv2z_all} c) which does not depend much on the approximation used. Furthermore, $\Delta E_{PV}$ has a very large variation between each of the different values of $ECM$, which are taken from each substituted molecule (being the atom A replaced by F, Cl, Br, I).

\begin{figure}[]
  \begin{minipage}[b]{0.5\linewidth}
    \centering
    \includegraphics[width=1.0\linewidth]{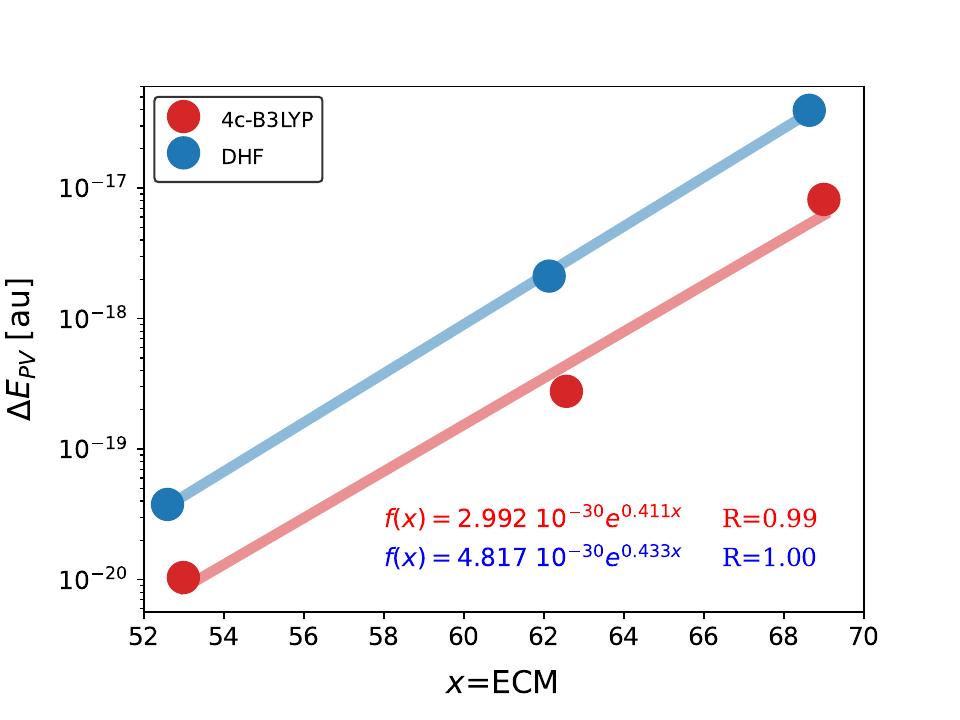}
  \end{minipage}
  \begin{minipage}[b]{0.5\linewidth}
    \centering
    \includegraphics[width=1.0\linewidth]{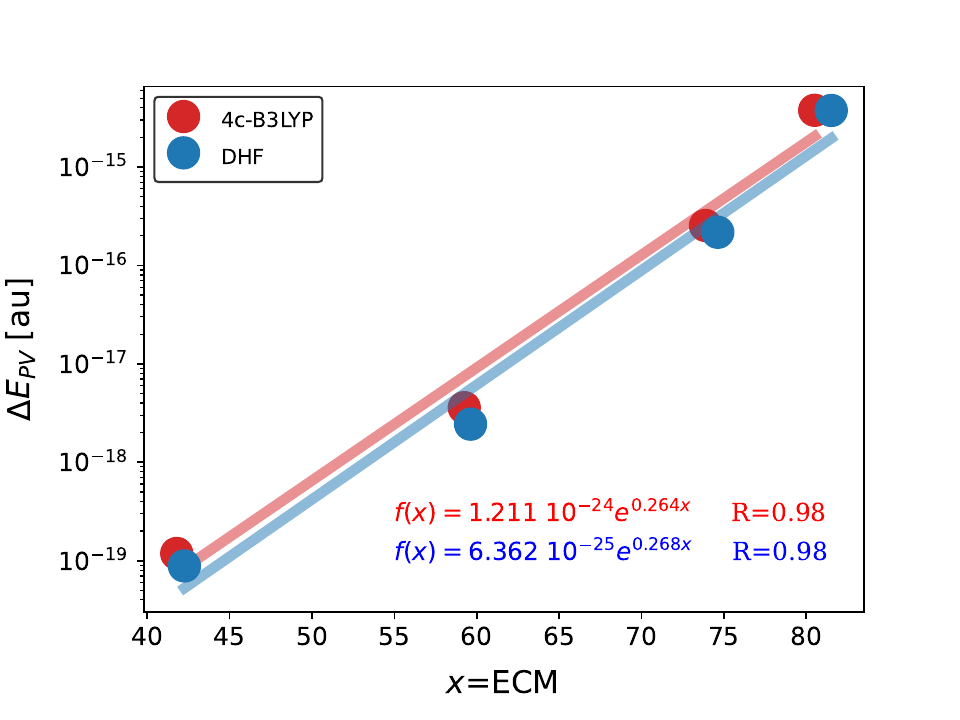}
  \end{minipage}
  \begin{minipage}[b]{0.5\linewidth}
    \centering
    \caption*{a)}
    \includegraphics[width=1.0\linewidth]
    {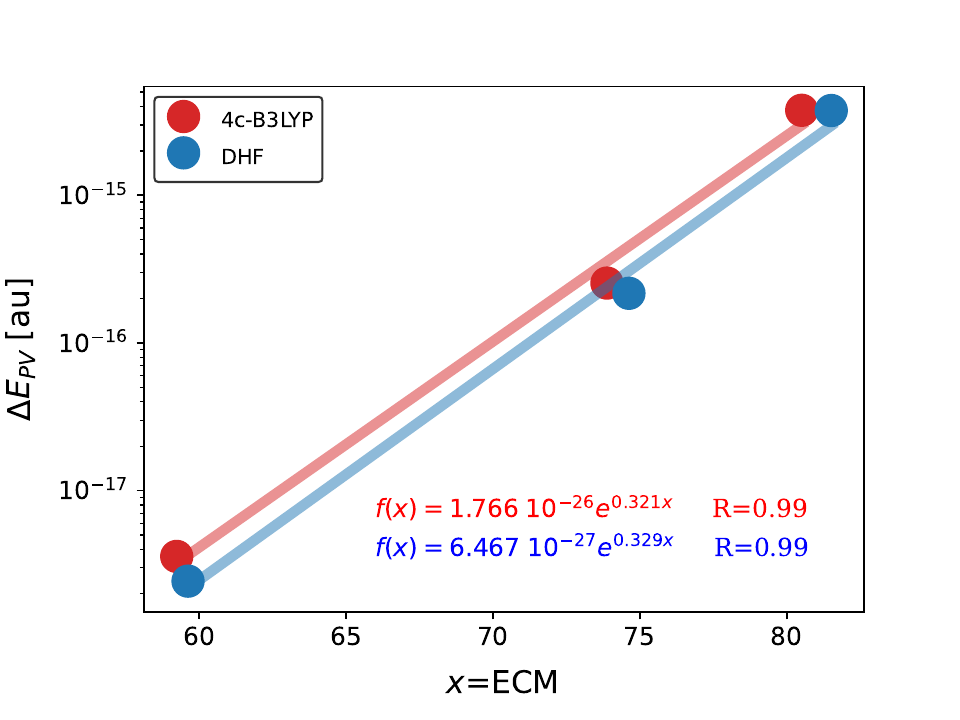}
    \caption*{c)}
  \end{minipage}
  \begin{minipage}[b]{0.5\linewidth}
    \centering
    \caption*{b)}
    \includegraphics[width=1.0\linewidth]{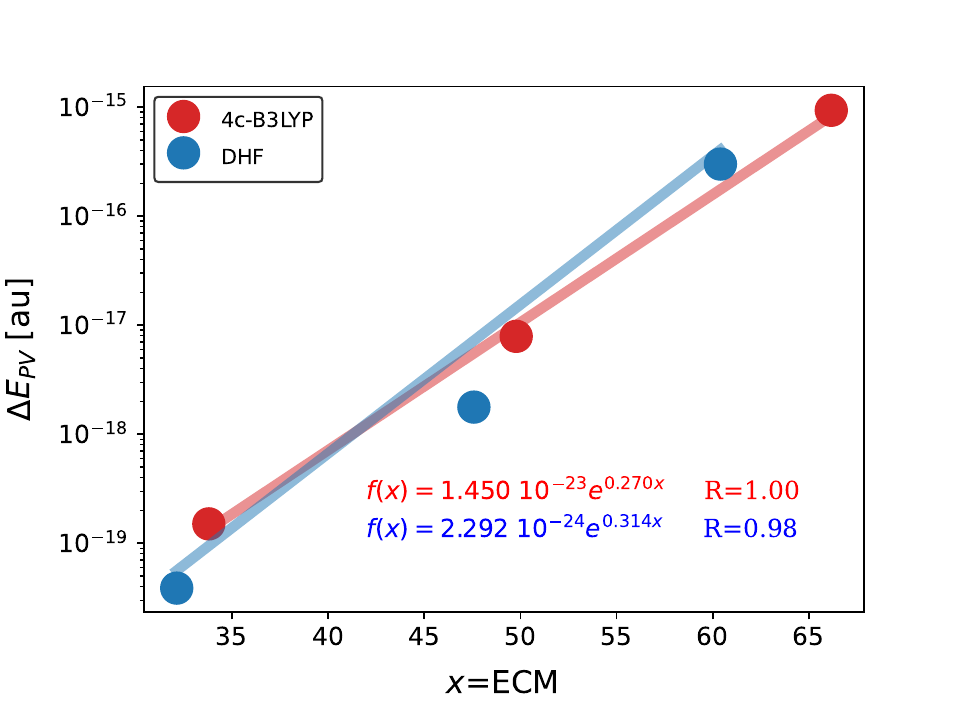}
    \caption*{d)}
  \end{minipage}
  \caption{
  Correlation between $\Delta E_{PV}$ and $ECM$ for the following molecules: CHEAR, a), alanine, b) and c) and glyceraldehyde, d). The substituted atoms whose atomic numbers were taken as a parameter are shown in the ESI file. Calculations were performed at DHF/dyall.cv2z and 4C-B3LYP/dyall.cv2z levels of theory.
  }
  \label{fig:REL_CHEAR_ala_gly_cv2z}
\end{figure}

In order to see whether these findings depends on the structure of the chiral molecule, we studied the CHEAR molecule whose plane of symmetry, the one that contains the nearest achiral symmetric structure, is far apart from the position of the substituted atoms. The results are shown in Fig. \ref{fig:REL_CHEAR_ala_gly_cv2z} a) where it is observed that the relation between $\Delta E_{PV}$ and $ECM$ holds the same. For this molecule we substituted the atom N by P, As and Sb (see the ESI file). Again, the electron correlation effects are within the same order of magnitude for each substituent. Note however that the CHEAR-type structure with Sb is not analyzed here since the molecular geometry becomes very distorted.

Let us now consider two completely different kind of geometrical structures: alanine and glyceraldehyde.
In Fig. \ref{fig:REL_CHEAR_ala_gly_cv2z} b) and
Fig. \ref{fig:REL_CHEAR_ala_gly_cv2z} c) we show the results for alanine. The two oxygen atoms are replaced by S, Se and Te (see Fig. 3 of ESI) being the Polonium excluded since for this substituent the molecule becomes highly distorted. There we observe a linear relation between $\Delta E_{PV}$ and $ECM$ for both kind of four-component calculations, DHF and 4C-B3LYP. As one may expect, there are no difference of slopes when electron-electron interactions are included at 4C-DFT level of approach.

For alanine, we found a small deviation from the positive linear correlation (see Fig. \ref{fig:REL_CHEAR_ala_gly_cv2z} b). This behavior likely comes from the fact that the contribution to $\Delta E_{PV}$, originated in the nitrogen atom, is similar to that arising from the oxygen atoms, thus giving two counterbalancing contributions. If the first point of the Fig. \ref{fig:REL_CHEAR_ala_gly_cv2z} b) (which arises by considering the oxygen as substituent atoms) is not considered, the $R^2$ value becomes larger than 0.99 (see Fig. \ref{fig:REL_CHEAR_ala_gly_cv2z} c)).

The last molecule analysed was glyceraldehyde. In this case, we substituted three oxygen atoms by sulfur and selenium (see Fig. 4 of the ESI). The results are presented in Fig. \ref{fig:REL_CHEAR_ala_gly_cv2z} d). Tellurium was not considered. We must stress here that when including the substituents of sulfur and selenium, there are significant structural distortions. However, also in this case, the relation between $\Delta E_{PV}$ and $ECM$ retain its exponential dependence. Furthermore, as occurs for alanine, the value of $\Delta E_{PV}$ are the highest obtained in this study.

\begin{figure}[]
  \begin{minipage}[b]{0.5\linewidth}
    \centering
    \includegraphics[width=1.0\linewidth]{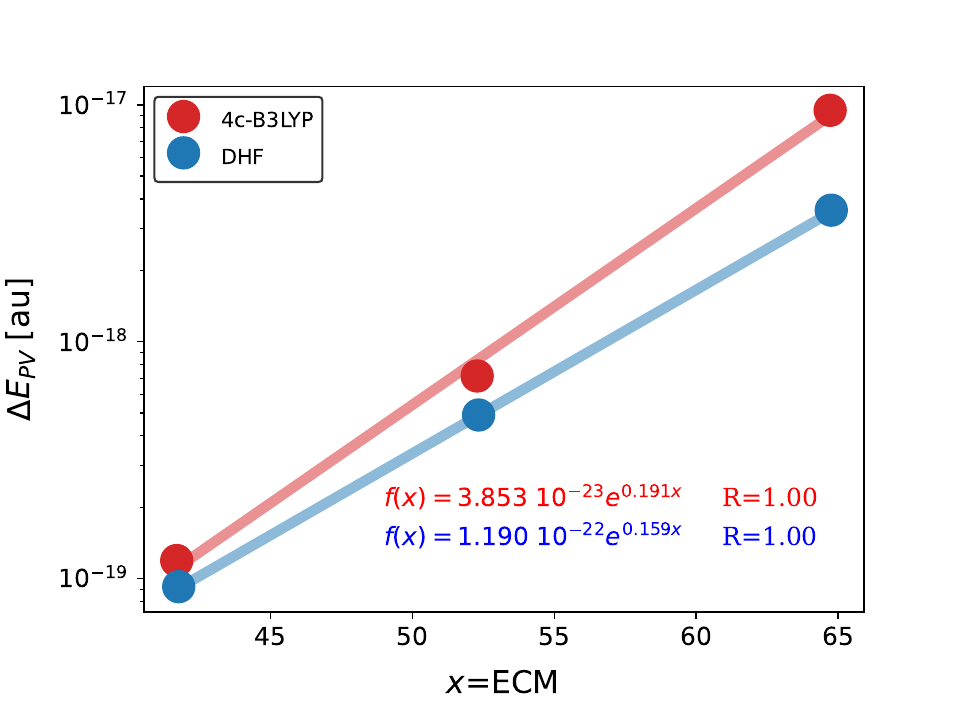}
  \end{minipage}
  \begin{minipage}[b]{0.5\linewidth}
    \centering
    \includegraphics[width=1.0\linewidth]{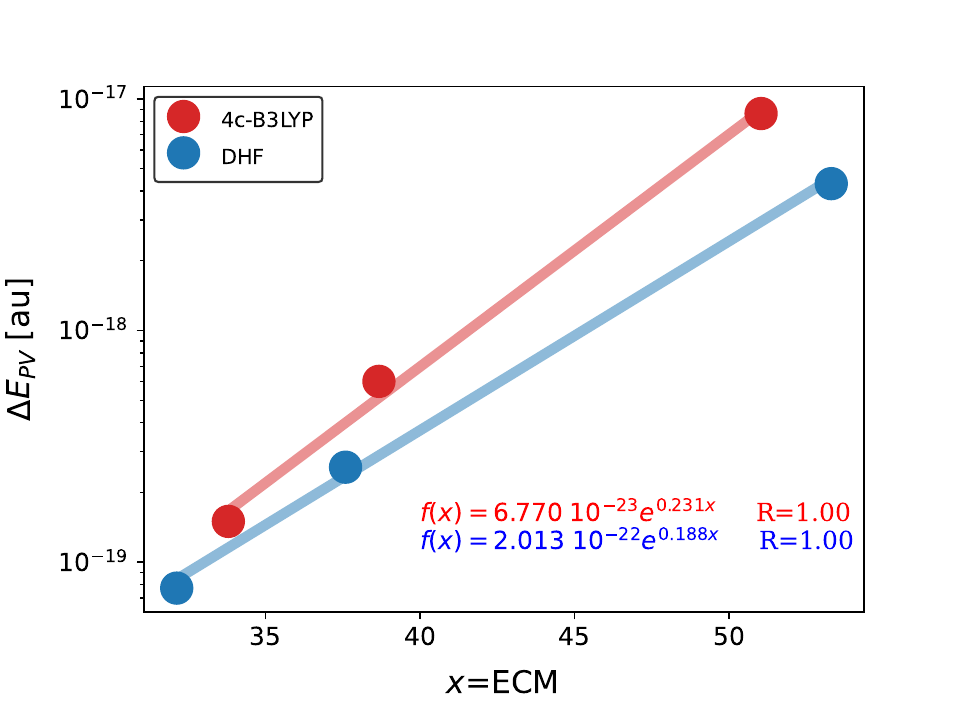}
  \end{minipage}
  \caption{
  Correlation between $\Delta E_{PV}$ and $ECM$ for the following molecules: alanine a), glyceraldehyde b). The substituted atoms whose atomic numbers were taken as a parameter are shown in the ESI file. Calculations were performed at DHF/dyall.cv2z and 4C-B3LYP/dyall.cv2z levels of theory.
  }
  \label{fig:REL_ala_gly_cv2z_1atom}
\end{figure}

It is nicely seen that, for alanine, the values of the $ECM$ are larger than its values for the other compounds when the substituent atoms belong to the same column of the Periodic Table. There is a similar behavior for $\Delta E_{PV}$. This fact is also observed for glyceraldehyde but in this last case the values of $ECM$ are smaller than in alanine.
This behavior follows what would be expected from the single-center theorem of Hegstrom-Rein-Sandars\cite{1980_Hegstrom}, though now in terms of $ECM$. As observed in Fig. \ref{fig:REL_ala_gly_cv2z_1atom}, when we systematically replace only one atom instead of two or three, as we did in alanine and glyceraldehyde, respectively, the values of $ECM$ become much smaller, as happens to $\Delta E_{PV}$ that is reduced in two orders of magnitude.

\section{Conclusions}
After quite a long term of research of the physical origin of the molecular chirality, still there are no experimental evidence about it, neither about the biological evolution of homochirality. On the other hand, highly accurate calculations of parity-violation (PV) effects in a selected set of chiral molecules have shown that these effects are so small that the experimental detection is extremely difficult. This suggests that the search for PV effects in molecules and solids should consider other alternatives, like computational experiments.

In line with this, quantum chemistry studies have considered recently the problem of quantifying the molecular chirality. For example, the electron chirality measure ($ECM$) provides a metric to quantify how far away a chiral molecule is from its nearest reference achiral partner, making use of the inner projection technique.

As another step forward of previous studies we have shown here that the molecular chirality can be related with the weak force between electrons and nucleons acting mainly within the nuclei. We found a linear correlation between the difference of the PV contribution to the electronic energy of two chiral enantiomers, $\Delta E_{PV}$ and the descriptor $ECM$. This also suggests that experimental investigations for detecting the likely parity violating molecular effects should start to consider molecules with as large $ECM$ values as possible.

Finally we provide a code for the calculation of $ECM$ in molecules within the relativistic four component framework.
\section{Supporting Information}
Schemes for the molecules under study and the selected substituent atoms. \\
PyECM implementation details.
\begin{acknowledgement}
The argentinian authors wants to thank the Institute for Modeling and Innovative Technologies (IMIT) of the National Research Council for Science and Technique (CONICET), and the Northeastern University of Argentina (UNNE) for their support and for providing access to the IMIT high performance computing cluster.
Support from CONICET by Grant No. PIP 1122 02101 00483 and FONCYT by Grant No. PICT-2021 I-A-00933 is greatly acknowledged.
Mr. Juan. J. Aucar would also like to thank acknowledges the hospitality by CNR-SPIN c/o Department
of Physical and Chemical Science at University of L {'}Aquila (Italy) during the visit. The work has been performed under the projects HPC-EUROPE3 (HPC17QOVMV, HPC174MQ8N), with the support of the EC Research Innovation Action under the H2020 Programme; in particular, the author gratefully acknowledges the support the computer resources and technical support provided by CINECA.
They also acknowledge the CINECA award under the ISCRA initiative, for the availability of high performance computing resources and support. A.S. acknowledges support from the Italian Ministry of Research under the  PRIN 2022 Grant No 2022F2K7J5 with title \textit{Two-dimensional chiral hybrid organic-inorganic perovskites for chiroptoelectronics} PE 3
funded by PNRR Mission 4 \textit{Istruzione e Ricerca} - Component C2 - Investimento 1.1, \textit{Fondo per il Programma Nazionale di Ricerca e Progetti di Rilevante Interesse Nazionale} PRIN 2022 - CUP B53D23004130006.
\end{acknowledgement}

\bibstyle{achemso}
\bibliography{biblio}

\end{document}